\documentclass[11pt,letterpaper]{article}
\pdfoutput=1
\usepackage{jcappub}

\usepackage{appendix}
\usepackage{array}
\newcolumntype{x}[1]{>{\centering\arraybackslash}p{#1}}

\usepackage{amsmath}
\usepackage{amsfonts}
\usepackage{amssymb}
\usepackage{eucal}               % Modifies style {\cal M} e \mathcal{M}. Old characters available with command \CMcal{M}.
\usepackage{mathrsfs}          % Cursive letter for Lagrangian and  matrix elements \mathscr{M}
\usepackage{slashed}

\usepackage{epsfig}
\usepackage{graphicx}
\usepackage{dcolumn}
\usepackage{enumerate}

\usepackage{color}
\usepackage{ulem}
\hypersetup{colorlinks=false}

\newcommand{\beq}{\begin{equation}}
\newcommand{\eeq}{\end{equation}}
\newcommand{\ud}{\text{d}}

\newcommand{\Eq}[1]{Eq.~\eqref{#1}}

\newcommand{\mDM}{m}
\newcommand{\ER}{E_\text{R}}
\newcommand{\Ed}{E'}

\newcommand{\vmin}{v_\text{min}}

\newcommand{\bol}[1]{\boldsymbol{#1}}
\newcommand{\bfv}{\bol{v}}
\newcommand{\eH}{\mathcal{H}}
\newcommand{\eR}{\mathcal{R}}

\title{Direct Detection of Light ``Ge-phobic" Exothermic Dark Matter}

\author[a]{Graciela B. Gelmini,}
\author[a]{Andreea Georgescu}
\author[a]{and Ji-Haeng Huh}

\affiliation[a]{Department of Physics and Astronomy, UCLA,\\
475 Portola Plaza, Los Angeles, CA 90095, USA}

\emailAdd{gelmini@physics.ucla.edu}
\emailAdd{a.georgescu@physics.ucla.edu}
\emailAdd{jhhuh@physics.ucla.edu}

\abstract{
We present comparisons of direct dark matter (DM) detection data for light WIMPs with exothermic scattering with nuclei (exoDM), both assuming the Standard Halo Model (SHM) and in a halo model -- independent manner.  Exothermic interactions favor light targets, thus reducing the importance of  upper limits derived from xenon targets, the most restrictive of which is at present  the  LUX  limit.  In our  SHM analysis the CDMS-II-Si and CoGeNT regions become allowed by these bounds,  however the recent SuperCDMS limit rejects both regions for exoDM with isospin-conserving couplings. An isospin-violating coupling of the exoDM, in particular one with a  neutron to proton coupling ratio  of $-0.8$ (which  we call ``Ge-phobic''), maximally reduces the DM coupling to germanium and allows the CDMS-II-Si region to become compatible with all bounds. This is also clearly shown in our halo-independent analysis.
}


\notoc

\begin{document}

\maketitle

\setcounter{page}{0}

\section{Introduction}
WIMPs,  weakly interacting massive particles, are the dark matter (DM) candidates most actively searched for.  Several potential signals for  ``light WIMPs'', i.e.\ WIMPs with mass around $1$--$10$ GeV/$c^2$, have appeared in four direct detection searches: DAMA~\cite{Bernabei:2003za, Bernabei:2008yi, Bernabei:2010mq} (here DAMA stands for both DAMA and  DAMA/LIBRA), CoGeNT~\cite{Aalseth:2010vx, Aalseth:2011wp, Aalseth:2012if,Aalseth:2014eft,Aalseth:2014jpa},  CRESST-II~\cite{Angloher:2011uu}, and  more recently 
CDMS-II-Si \cite{Agnese:2013rvf}, either as an unexplained  excess of events (in CoGeNT, CRESST-II and CDMS-II-Si) or as an annual modulation of the rate as expected for a DM signal (DAMA and CoGeNT).
It was first shown in 2004 \cite{Gelmini:2004gm}  that  light WIMPs with spin-independent isospin-conserving interactions with nuclei  could  simultaneously provide a viable DM interpretation of the DAMA annual  modulation~\cite{Bernabei:2003za} and be compatible with all negative searches at the time, assuming   the Standard Halo Model (SHM) for the dark halo of our galaxy.
The interest in these candidates intensified as new hints of light WIMPs appeared,  first  in the  DAMA 2008 data \cite{Bernabei:2008yi} (see e.g.~\cite{Savage:2008er}) and then  in the data of CoGeNT, CRESST-II and CDMS-II-Si.  In the following we do not include the regions of interest due to CRESST-II because of the difficulty we found in the analysis of their data, but we include  all the other regions, and the most relevant limits derived from  direct DM searches  with null results by  the LUX~\cite{Akerib:2013tjd}, XENON10~\cite{Angle:2011th}, CDMSlite \cite{Agnese:2013jaa}, SuperCDMS~\cite{Agnese:2014aze} and SIMPLE~\cite{Felizardo:2011uw} collaborations (with SIMPLE relevant only for isospin-violating~\cite{Kurylov:2003ra, Feng:2011vu} interactions).

As pointed out recently in previous direct DM detection data analysis~\cite{Frandsen:2013cna, McCullough:2013jma, Fox:2013pia, Frandsen:2014ima}, of particular interest for the compatibility of the potential CDMS-II-Si  and CoGeNT signals with all present limits is the possibility of having DM with inelastic exothermic collisions  with nuclei, originally called  ``exciting DM"~\cite{Finkbeiner:2007kk} in the context of indirect DM detection and later  ``exothermic DM" (exoDM)~\cite{Batell:2009vb, Graham:2010ca} in the context of direct dark matter detection. Having a complicated ``dark sector"~\cite{ArkaniHamed:2008qn, Batell:2009vb, Graham:2010ca,  Essig:2010ye}  with neutral particles of slightly different masses leads naturally to the idea of having  two different states constituting the DM at present, the lightest being stable and the heaviest metastable. It can then happen that the heaviest may down-scatter off nuclei, but the scattering of each state to itself is suppressed  or impossible because of the DM couplings to the mediator of the interaction, and the up-scattering of the lightest  state  is kinematically forbidden (as we will see below, the required speeds for the models we consider would be above $1000$ km/s  and these high WIMP speeds are not available in the halo of our galaxy). This type of DM favors lighter targets with respect to heavier ones, thus it suppresses the limits derived from experiments using xenon, which provide otherwise some of the most restrictive limits at present.

Here we consider exoDM  as a potential explanation of the signals found in direct detection data, both assuming the SHM and in a halo-independent manner.  The halo-independent analysis  of direct DM detection data was proposed and later used in different forms in Refs.~\cite{Fox:2010bz, Frandsen:2011gi, HerreroGarcia:2011aa, Gondolo:2012rs, HerreroGarcia:2012fu, DelNobile:2013cta,  Bozorgnia:2013hsa, DelNobile:2013cva, DelNobile:2013gba, DelNobile:2014eta}. The method was generalized  in \cite{DelNobile:2013cva} to be applied to WIMP-nucleus scattering cross sections with any type of speed dependency~\cite{DelNobile:2014eta}. This method consists in mapping the rate measurements and bounds onto $\vmin$ space, where $\vmin$ is the minimum WIMP speed necessary to impart a recoil energy $\ER$ to the target nucleus. This method allows to factor out a common function of $\vmin$ which gives the dependency of the rate on the DM velocity distribution, and use this as a detector-independent variable. Since $\vmin$ is also a detector-independent quantity (while $\ER$ depends on the target mass), outcomes from different direct detection experiments can be directly compared in $\vmin$ space. 
 
The data analysis in this paper is the same as in~\cite{DelNobile:2014eta}, except for the binning of the CoGeNT data for our halo-independent analysis (here we took two bins, 0.5--2.0 KeVee and 2.0--4.5 keVee, the same as in~\cite{Aalseth:2014eft}). The SHM is also the same as in~\cite{DelNobile:2014eta} except that here we take the escape speed from our galaxy at the location of the Solar System to be the median-likelihood value of the most recent  Radial Velocity Experiment (RAVE) 2013 results~\cite{Piffl:2013mla}, 533$^{+54}_{-41}$ km/s (instead of the more frequent choice of 544 km/s --- the median-likelihood value of the RAVE 2006 results~\cite{Smith:2006ym}). However, this change of escape speed has negligible effects in  all the plots we show.

\section{Inelastic exothermic scattering}

In certain particle models, a DM particle of mass $m$ may collide inelastically with  a target nucleus  producing a different state with mass $m' = m + \delta$ while the elastic scattering is either forbidden or suppressed~\cite{TuckerSmith:2001hy}. 
In inelastic scattering, the minimum speed  $\vmin$ the DM particle must have in order to impart a nuclear recoil energy $\ER$ to a nucleus of mass $m_T$ depends on the mass splitting $\delta$. For $\mu_T\left| \delta \right|/m^2 \ll 1$, $\vmin$  is
\beq
\vmin = \frac{1}{\sqrt{2 m_T \ER}} \left| \frac{m_T \ER}{\mu_T} + \delta \right| ,
\eeq
where $\delta$ can be either positive for an endothermic scattering~\cite{TuckerSmith:2001hy} or negative for an exothermic scattering~\cite{Finkbeiner:2007kk,Batell:2009vb,Graham:2010ca} (with $\delta = 0$ for elastic scattering). Here $\mu_T$ is the WIMP-nucleus reduced mass. The same equation relates the speed of a DM particle $v$ with the maximum, and in this case also minimum, recoil energy the particle can impart to a nucleus. Inverting this equation one finds the maximum and minimum recoil energies for a fixed DM  particle speed $v$: $\ER^-(v) < \ER < \ER^+(v)$, with
\beq
\ER^\pm(v) =
\frac{\mu_T^2 v^2}{2 m_T} \left( 1 \pm \sqrt{1 - \frac{2 \delta}{\mu_T v^2}} \right)^2 .
\eeq

\begin{figure}[t!]
\centering
\includegraphics[width=0.49\textwidth]{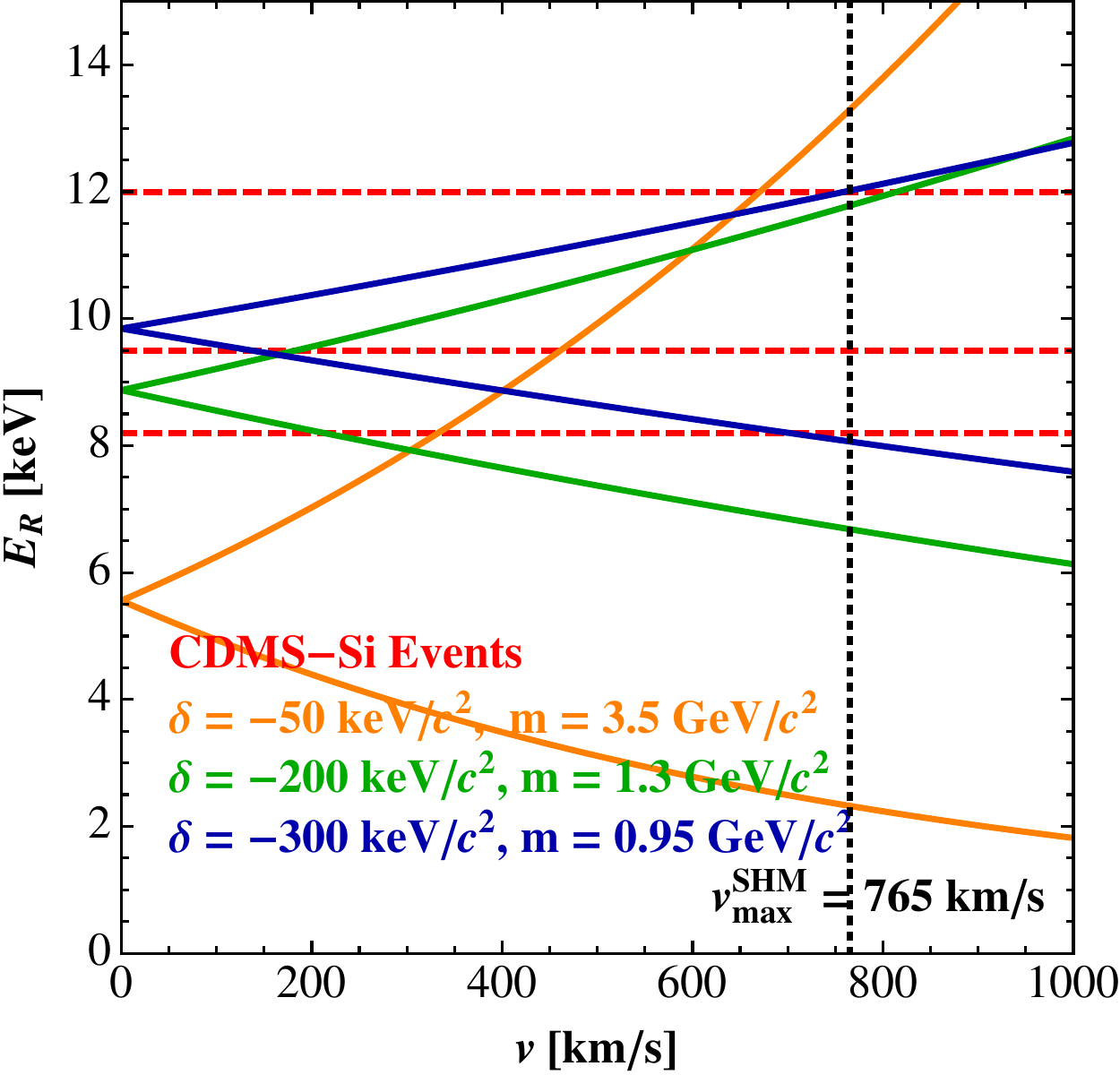}
\caption{\label{fig:ErVsV}
Recoil energy range in Si as a function of  the WIMP speed $v$ with respect to the Earth for different values of the WIMP mass $m$ and mass splitting $\delta$, compared with the energies  of the three events observed by CDMS-II-Si (horizontal lines), assuming that the observed and recoil energies coincide. We see that for negative $\delta$ values with $\left| \delta \right| > 300$ keV/c$^2$ not all three events can be contained within the possible DM recoil energy range in the SHM (the halo model determines the maximum possible value of $v$,  which here is 765 km/s).
}
\end{figure}

Fig.~\ref{fig:ErVsV} shows these two functions for a  silicon target for several WIMP masses  $m$ and  negative values of $\delta$. For a particular recoil energy $\ER$ only the speeds to the right of the $\vmin(\ER)$ line are allowed, while for a fixed speed $v$ only the recoil energies in between the two lines $\ER^- (v)$ and $\ER^+ (v)$ are allowed. The maximum possible value of $v$  is the sum of the escape speed and the speed of the Earth with respect to the Galaxy, which for our SHM is 765 km/s. The maximum and minimum recoil energies $\ER^+ (v)$ and $\ER^- (v)$ meet with a value $\ER^+ (v_\delta)=\ER^- (v_\delta)=  \mu_T \left| \delta \right|/ m_T \equiv E_ \delta$, where $v_\delta$ is the minimum  possible value of $\vmin $. For $\delta > 0$, $v_\delta = \sqrt{2 \delta / \mu_T}$, and for $\delta \leqslant 0$,  $v_\delta = 0$. For positive $\delta$ we get $v_\delta \simeq 1000 \sqrt{(\delta/50\text{ keV})(10\text{ GeV}/m)}$~km/s if we assume that the WIMP is much lighter than the target nucleus so that  $\mu_T \simeq m$ (a good approximation in all the cases we consider here). This value of $v_\delta$ is too large for WIMPs in the dark halo of our galaxy, thus inelastic endothermic collisions cannot happen for a $\left|\delta\right|$ of 50 keV$/c^2$ or larger. In an exothermic scattering, the energy of the recoiling nucleus is peaked around $E_ \delta$, which is proportional to the splitting between the dark matter states and is inversely proportional to the nuclear mass. Consequently, the nuclear recoils caused by exoDM are more visible in experiments with light nuclei and low thresholds~\cite{Graham:2010ca}. 

In the same plot, we compare the allowed recoil energy values in silicon with the energies of the three events observed by CDMS-II-Si (horizontal lines). We took the recoil energies to coincide with the observed energies (i.e.\ perfect energy resolution), which is not a bad approximation for CDMS-II-Si. The values of $m$ and $\delta$ are those corresponding to some of the best fit points allowed by all upper limits (and coincide with the values of the halo-independent analysis we show later in the paper). We see that for negative $\delta$ values such that $\left| \delta \right| > 300$ keV$/c^2$ not all three events can be contained within the possible recoil energy range for the SHM, i.e.\ either the largest energy event (or events) or the lowest energy event (or events) must be due to background. However, even for smaller negative $\delta$ values, such as $\delta=-200$ keV$/c^2$, we find that, in the SHM, best fit regions are obtained when the highest-energy of the three CDMS-II-Si observed events is considered background.

The  event rate in a detected energy interval $[\Ed_1, \Ed_2]$ is 
\begin{multline}
\label{R-inelastic}
R_{[\Ed_1, \Ed_2]}(t) =
\frac{\rho}{\mDM} \sum_T \frac{C_T}{m_T} \int_0^\infty \ud \ER \, \int_{v \geqslant v_\text{min}(\ER)} \hspace{-18pt} \ud^3 v \, f(\bfv, t) \, v \, \frac{\ud \sigma_T}{\ud \ER}(\ER, \bfv)
\\
\times
 \int_{\Ed_1}^{\Ed_2} \ud\Ed \, \epsilon(\Ed) G_T(\ER, \Ed)
.
\end{multline}
Here the sum runs over all the different target nuclides $T$, $C_T$ is the mass fraction of the target $T$, $\rho$ is the local DM density, $f(\bfv, t)$ is the DM velocity distribution in the reference frame of the Earth, and $\epsilon(\Ed)$ is the experimental acceptance. $G_T(\ER, \Ed)$ is the detector target-dependent resolution function expressing the probability that a recoil energy $\ER$ is measured as $\Ed$, and incorporates the mean value $\langle \Ed \rangle = Q_T(\ER) \ER$ (with $Q_T$ the quenching factor) and the detector energy resolution.

For our halo-independent analysis we follow the procedure presented in the Appendix A  of~\cite{DelNobile:2013cva}. Changing the order of the integrations in ${\bf v}$ and $\ER$ in \Eq{R-inelastic}, we have
\begin{multline}\label{R6}
R_{[\Ed_1, \Ed_2]}(t) =
\frac{\rho \sigma_{\rm ref}}{\mDM} \int_{v \geqslant v_\delta} \ud^3 v \, \frac{f(\bfv, t)}{v}
\sum_T \frac{C_T}{m_T} \int_{\ER^-(v)}^{\ER^+(v)} \ud \ER \, \frac{v^2}{\sigma_{\rm ref}} \frac{\ud \sigma_T}{\ud \ER}(\ER, \bfv)
\\
\times
 \int_{\Ed_1}^{\Ed_2} \ud\Ed \, \epsilon(\Ed) G_T(\ER, \Ed) ,
\end{multline}
where $v_\delta$ is the minimum value $\vmin $ can take, with $v_\delta = \sqrt{2 \delta / \mu_T}$ for $\delta > 0$ and $v_\delta = 0$ for $\delta \leqslant 0$. $\sigma_{\rm ref}$ is a reference cross section, which in this paper will be the WIMP-proton cross section  $\sigma_p$ (see below). In compact form, \Eq{R6} reads
\beq
R_{[\Ed_1, \Ed_2]}(t) =  \int_{v \geqslant v_\delta} \ud^3 v \, \frac{\tilde{f}(\bfv, t)}{v} \, \eH_{[\Ed_1, \Ed_2]}(\bfv) ,
\eeq
where 
\beq
\tilde{f}(\bfv, t) \equiv \frac{\rho \sigma_{\rm ref}}{\mDM} \, f(\bfv, t)
\eeq
and
\begin{multline}
\eH_{[\Ed_1, \Ed_2]}(\bfv) \equiv
\sum_T \frac{C_T}{m_T} \int_{\ER^-(v)}^{\ER^+(v)} \ud \ER \, \frac{v^2}{\sigma_{\rm ref}} \frac{\ud \sigma_T}{\ud \ER}(\ER, \bfv)
 \int_{\Ed_1}^{\Ed_2} \ud\Ed \, \epsilon(\Ed) G_T(\ER, \Ed) .
\end{multline}

In almost all cases of interest,  the  differential cross sections, and thus the functions $\eH_{[\Ed_1, \Ed_2]}$, depend only on the speed $v=|\bfv|$ and not on the whole velocity vector. This is true if the DM flux and the target nuclei are unpolarized and the detection efficiency is isotropic throughout the detector, which is the most common case. With this restriction, we define the response function  as the derivative of the ``integrated response function" $\eH_{[\Ed_1, \Ed_2]}(v)$
\beq
\eR_{[\Ed_1, \Ed_2]}(v) \equiv \frac{\partial \eH_{[\Ed_1, \Ed_2]}(v)}{\partial v}.
\label{eq:RT}
\eeq
The interaction rate is now (see~\cite{DelNobile:2013cva} for details)
\beq
\label{R4}
R_{[\Ed_1, \Ed_2]}(t) =  \int_ {v_\delta}^\infty \ud v \, \tilde{\eta}(v, t) \,  \eR_{[\Ed_1, \Ed_2]}(v) ,
\eeq
and the velocity integral $\tilde{\eta}$ is defined as
\beq
\label{eta0}
\tilde{\eta}(\vmin, t) \equiv \frac{\rho {\sigma_{\rm ref}}}{\mDM} \int_{v \geqslant \vmin} \ud^3 v \, \frac{f(\bfv, t)}{v} \equiv \int_{\vmin}^\infty \ud^3 v \, \frac{\tilde{f}(\bfv, t)}{v} .
\eeq

Due to the revolution of the Earth around the Sun,  the velocity integral  $\tilde{\eta}(\vmin,t)$ has an annual modulation generally well approximated by the first two terms of a harmonic series,
\beq
\label{etat}
\tilde{\eta}(\vmin, t) \simeq \tilde{\eta}^0(\vmin) + \tilde{\eta}^1(\vmin) \cos\!\left[ \omega (t - t_0) \right],
\eeq
where $t_0$ is the time corresponding to the maximum of the signal and $\omega = 2 \pi/$yr. The unmodulated and modulated components $\tilde{\eta}^0$ and $\tilde{\eta}^1$ enter respectively in the definition of unmodulated and modulated parts of the rate. We can then proceed to find the  averages of  $\tilde{\eta}^0$  and $\tilde{\eta}^1$ (the ``crosses" in our halo-independent plots) and  upper limits on them, as functions of $\vmin$, as explained in detail in~\cite{DelNobile:2013cva} (for the data analysis see~\cite{DelNobile:2014eta} and references therein).

With respect to the  WIMP-nucleus cross section and values of $m$ and $\delta$, in this paper we proceed in a phenomenological manner, without referring to particular DM particle models (although models with the required WIMP masses and  mass splittings have been proposed in~\cite{Frandsen:2014ima} - see also references therein).
We use the usual contact spin-independent (SI)  interaction cross section, which applies also to exoDM~\cite{Graham:2010ca}
\beq
\frac{\ud \sigma_T}{\ud \ER} = \sigma_T^{\rm SI}(\ER) \frac{m_T}{2 \mu_T^2 v^2} ,
\eeq
with
\beq\label{SIcrossection}
\sigma_T^{\rm SI}(\ER) = \sigma_p \frac{\mu_T^2}{\mu_p^2} [ Z_T + (A_T - Z_T)(f_n / f_p) ]^2 F_{{\rm SI}, T}^2(\ER) .
\eeq
Here $Z_T$ and $A_T$ are respectively the atomic and mass number of the target nuclide $T$, $F_{{\rm SI}, T}(\ER)$ is the nuclear SI form factor (which we take to be the Helm form factor~\cite{Helm:1956zz}), $f_n$ and $f_p$ are the effective DM couplings to neutrons and protons, $\sigma_p$ is the WIMP-proton cross section and $\mu_p$ is the DM-proton reduced mass. Thus, in this paper the reference cross section in the equations above is ${\sigma_{\rm ref}}= \sigma_p$.

The isospin-conserving choice $f_n = f_p$ is  usually assumed by the experimental collaborations. However, the value of $f_n/f_p$ that minimizes the coupling $\sum_T[1 + (f_n/f_p) (A_T-Z_T)/Z_T]^2 (C_T/m_T)$ for a particular target element, where the sum runs over its isotopes, is also possible. The isospin-violating  choice  $f_n / f_p = -0.7$~\cite{Kurylov:2003ra, Feng:2011vu} produces the maximum  cancellation of the WIMP coupling to xenon,  suppressing  very effectively the interaction cross section for this target. In our case, the exothermic character of the DM interactions weakens the xenon-based limits for large enough mass splitting. Thus we consider the value   $f_n / f_p = -0.8$, which suppresses most efficiently the WIMP coupling with a germanium target. This  ad-hoc choice, which we could call ``Ge-phobic", weakens the SuperCDMS limits maximally and it is equally motivated (or not motivated) as the ``Xe-phobic" -0.7 choice.

\begin{figure}[t]
\centering
\includegraphics[width=0.49\textwidth]{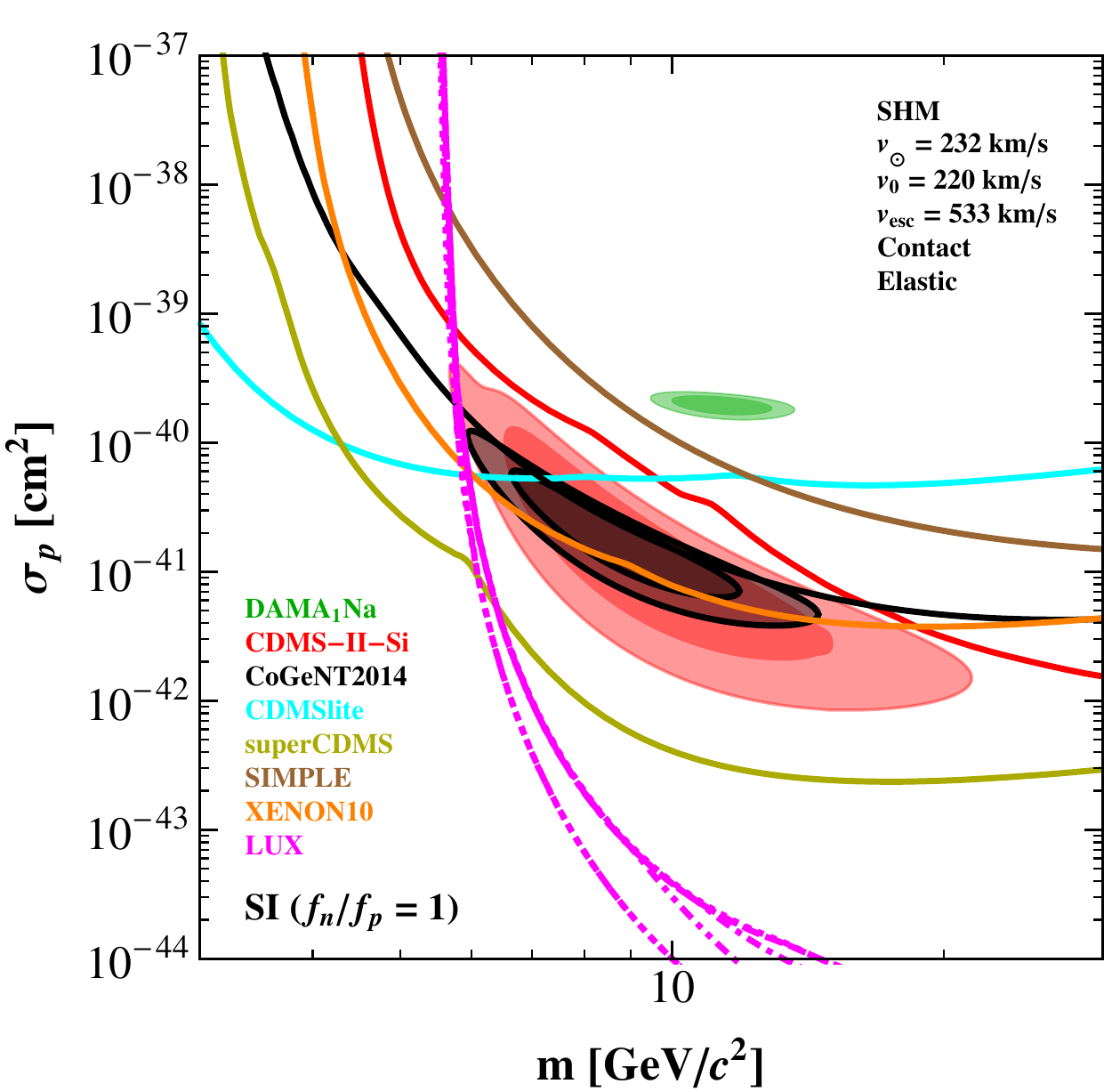}
\includegraphics[width=0.49\textwidth]{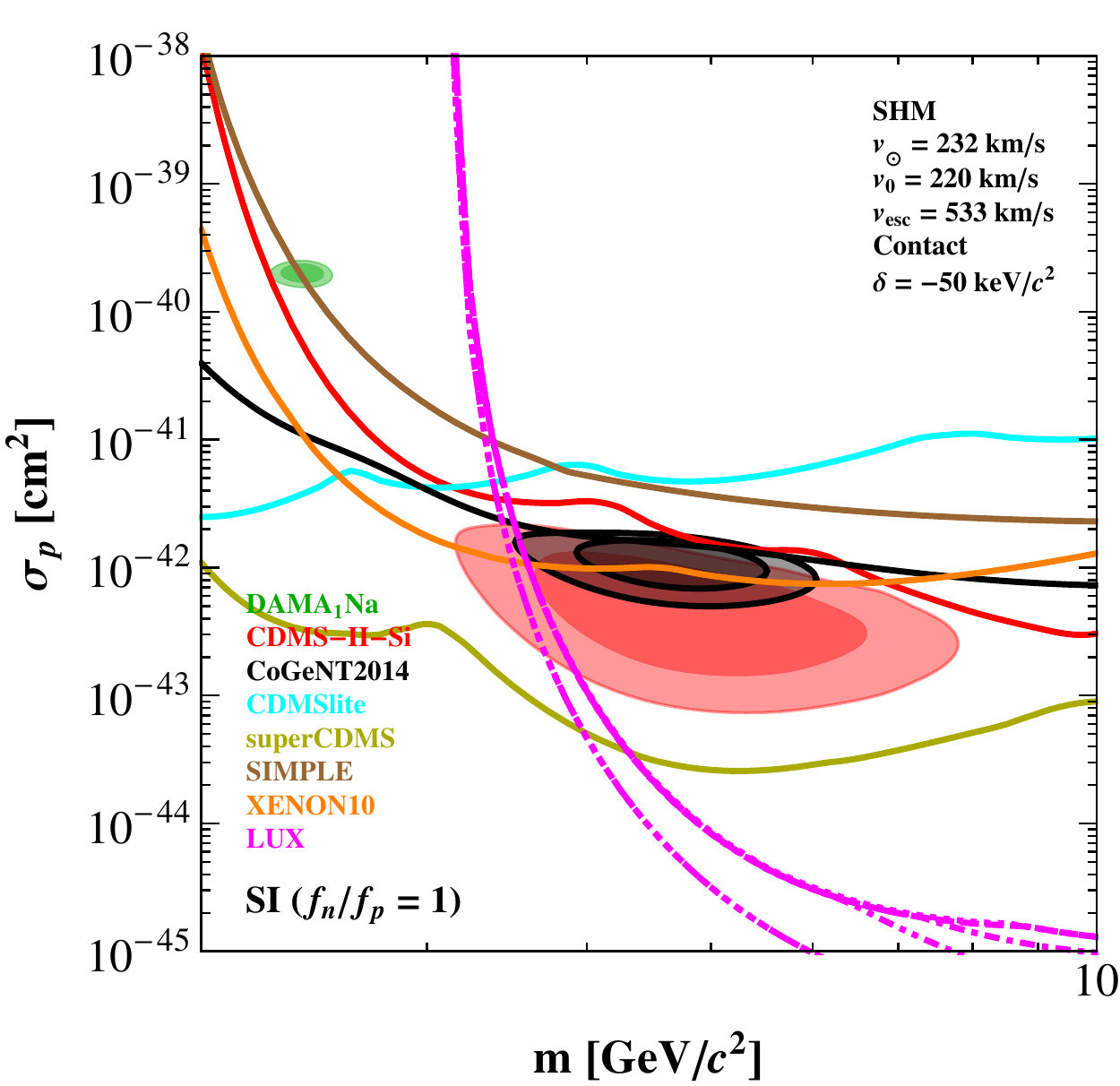}
\caption{\label{fig:fn1elastic50}
$90\%$ CL bounds and $68\%$ and $90\%$ CL allowed regions in the WIMP-proton cross section $\sigma_p$ vs WIMP mass plane, assuming the SHM, for spin-independent isospin-conserving interactions  for a) (left) elastic scattering ($\delta=0$) and b) (right)  inelastic exothermic scattering with $\delta=-50$ keV$/c^2$. The DAMA region is due to scattering off sodium (with $Q_{\rm Na} = 0.30$). 
}
\end{figure}
\begin{figure}[h]
\centering
\includegraphics[width=0.49\textwidth]{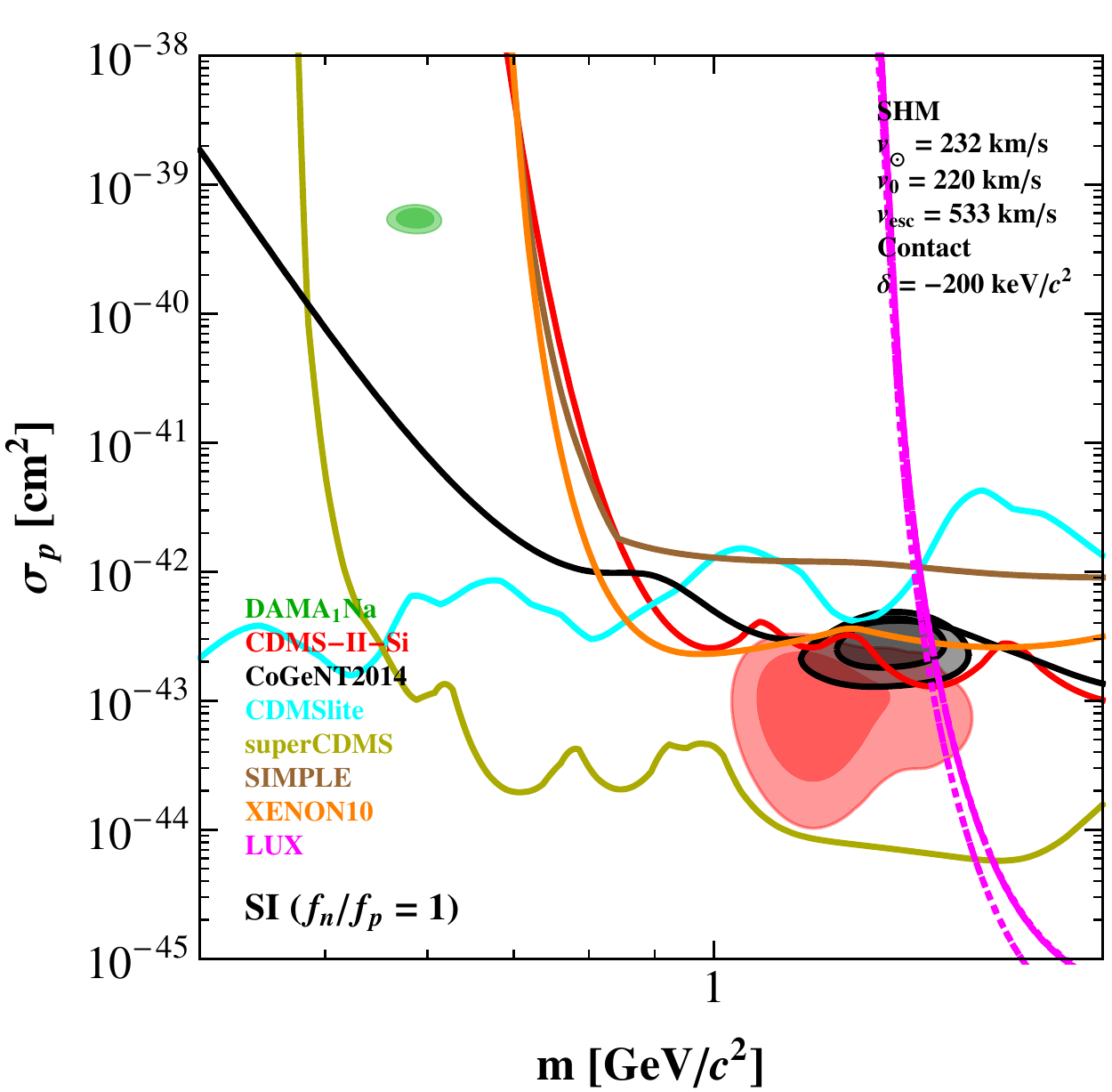}
\includegraphics[width=0.49\textwidth]{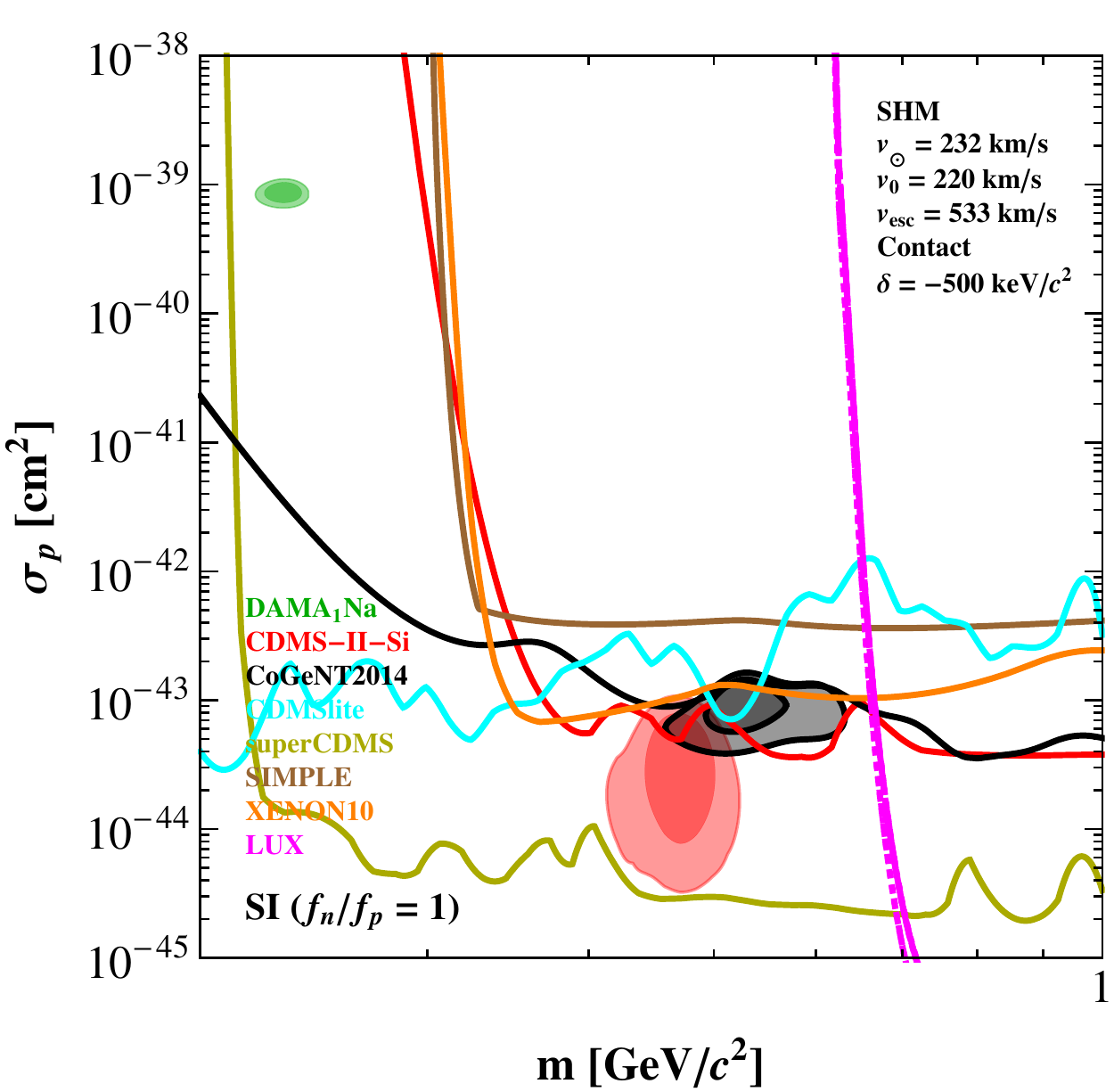}
\caption{\label{fig:fn1delta200500}
Same as Fig.~\ref{fig:fn1elastic50} but for  a) (left) $\delta=-200$ keV$/c^2$ and b) (right) $\delta=-500$ keV$/c^2$.
}
\end{figure}

\section{SHM data comparison}

 Figs.~\ref{fig:fn1elastic50} and \ref{fig:fn1delta200500}  show the $90\%$ confidence level (CL) bounds and $68\%$ and $90\%$ CL allowed regions for DAMA, CoGeNT $2014$ and CDMS-II-Si in the WIMP-proton cross section $\sigma_p$ vs WIMP mass $m$ plane, assuming the SHM,  for  spin-independent isospin-conserving interactions, and for elastic and inelastic scattering with $\delta = -50, -200$ and $ -500$ keV/c$^2$ (in Figs.~\ref{fig:fn1elastic50}.a, \ref{fig:fn1elastic50}.b, \ref{fig:fn1delta200500}.a and \ref{fig:fn1delta200500}.b respectively). As the mass difference $\delta$ between the  DM mass eigenstates increases, it becomes progressively more difficult to insure that the lifetime of the metastable DM state is larger than the lifetime of the Universe. Looking at Eq. 9 of  \cite{Frandsen:2014ima} (see also~\cite{Batell:2009vb}) it seems that the values we consider are still safe in this respect.

The irregular shape of the limits, more noticeable for larger negative $\delta$ values, is due to the rapid change of the interval chosen as the maximum gap since the narrow allowed energy range changes rapidly with $m$ (see Fig.~\ref{fig:ErVsV}). 

For CDMS-II-Si we found that, already for $\delta = -200$ keV/c$^2$, in the lower mass part of the best fit region the highest-energy event of the three observed events must be background (since a DM interaction would be kinematically forbidden; see Fig.~1). For $\delta = -500$ keV/c$^2$, in the lower mass part of the allowed CDMS-II-Si region only the lowest-energy event is due to DM, while in the higher mass part only the two lowest-energy events are due to DM.  This is not a problem in our statistical analysis because we have included both the signal and background contributions in the Extended Likelihood function~\cite{DelNobile:2014eta}. 

Notice how the DAMA region moves progressively to lower WIMP mass values with respect to the CoGeNT and CDMS-II-Si regions, as the negative $\delta$ value increases. This is because the signal in DAMA is the annual modulation of the rate, and the observed phase of the modulation require WIMP speeds larger than approximately 200 km/s in the SHM (for lower speeds the modulation amplitude changes sign, i.e. the times of maximum and minimum rate are reversed). With exothermic interactions even WIMPs with very low speeds could have energies above the experimental threshold of DAMA, unless the  WIMP mass is sufficiently small. The CoGeNT and CDMS-II-Si regions are derived from unmodulated rate measurements instead.

Figs.~\ref{fig:fn1elastic50} and \ref{fig:fn1delta200500} show that, when assuming the SHM, considering exoDM per se does not bring about compatibility between the potential signal regions and the upper limits in the $m$-$\sigma_p$ plane. The exothermic scattering is effective in weakening the xenon-based limits (the most important of which is LUX), but does little to suppress the germanium-based SuperCDMS limit which remains very restrictive because of the very low energy threshold of the experiment (1.6 keVnr).

\begin{figure}[t]
\centering
\includegraphics[width=0.49\textwidth]{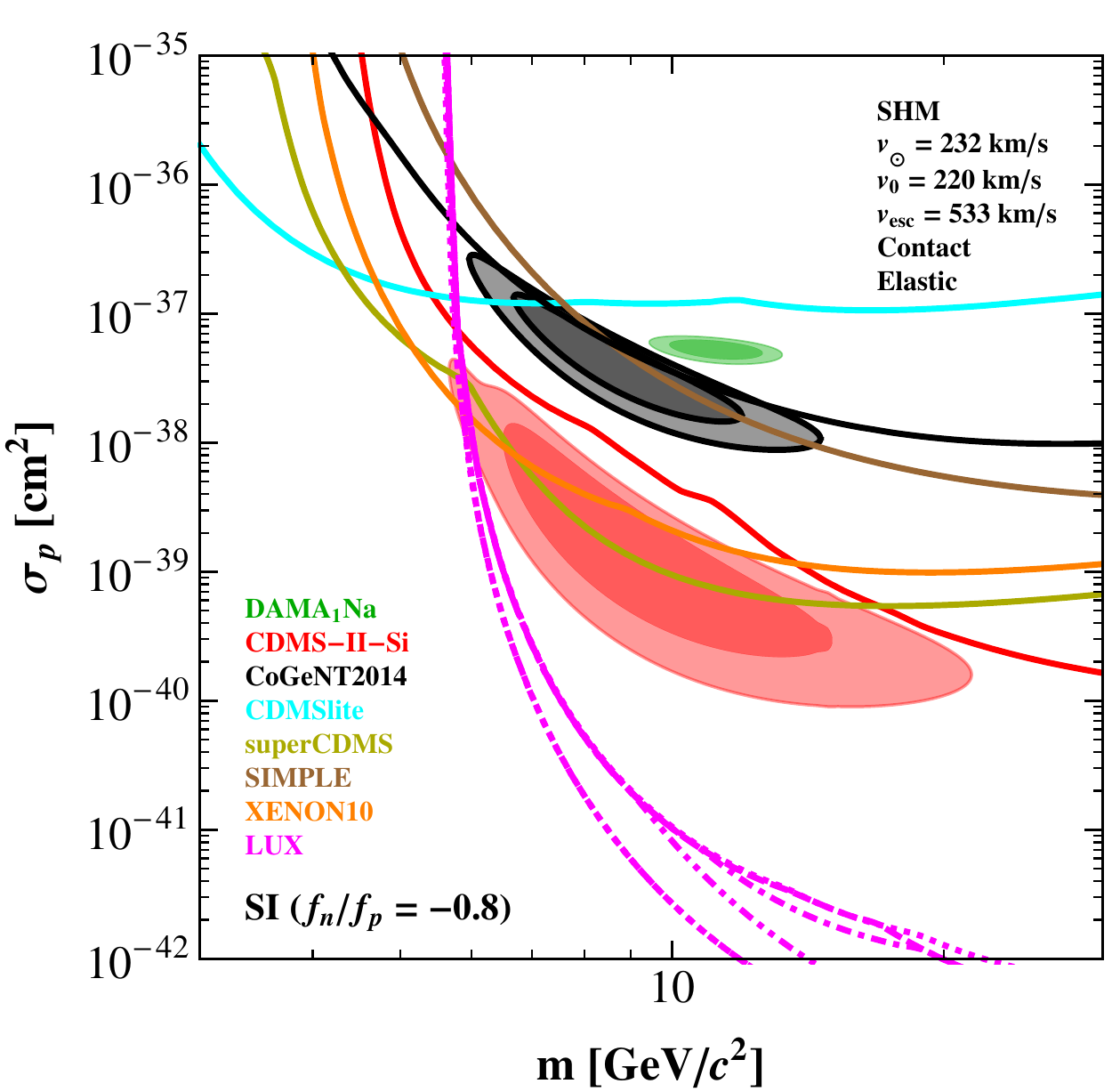}
\caption{\label{fig:fn08elastic}
$90\%$ CL bounds and $68\%$ and $90\%$ CL allowed regions in the WIMP-proton cross section $\sigma_p$ vs WIMP mass plane, assuming the SHM, for the spin-independent isospin-violating interactions with $f_n/f_p =-0.8$  and elastic scattering ($\delta=0$).
}
\end{figure}

\begin{figure}[t]
\centering
\includegraphics[width=0.49\textwidth]{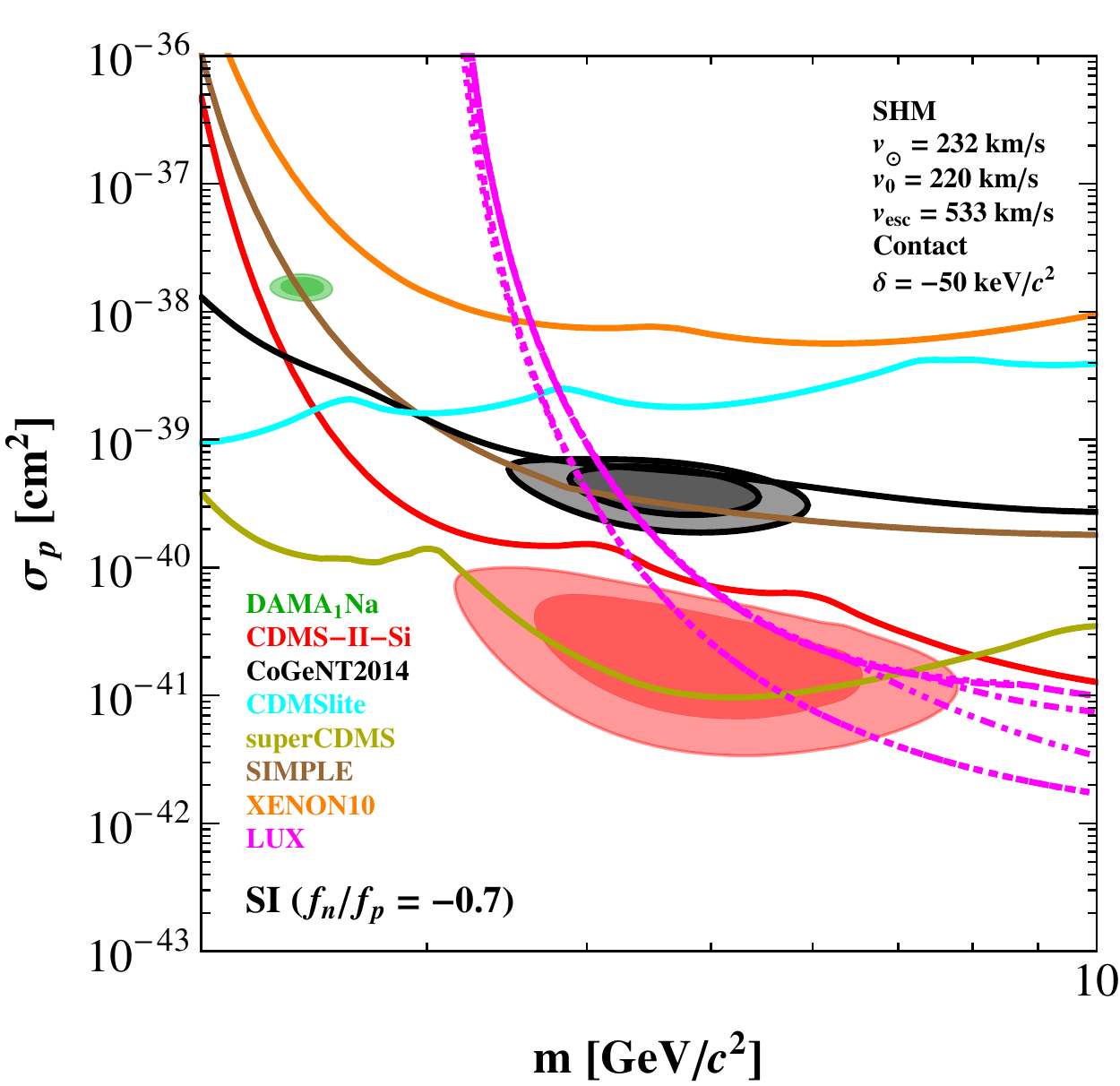}
\includegraphics[width=0.49\textwidth]{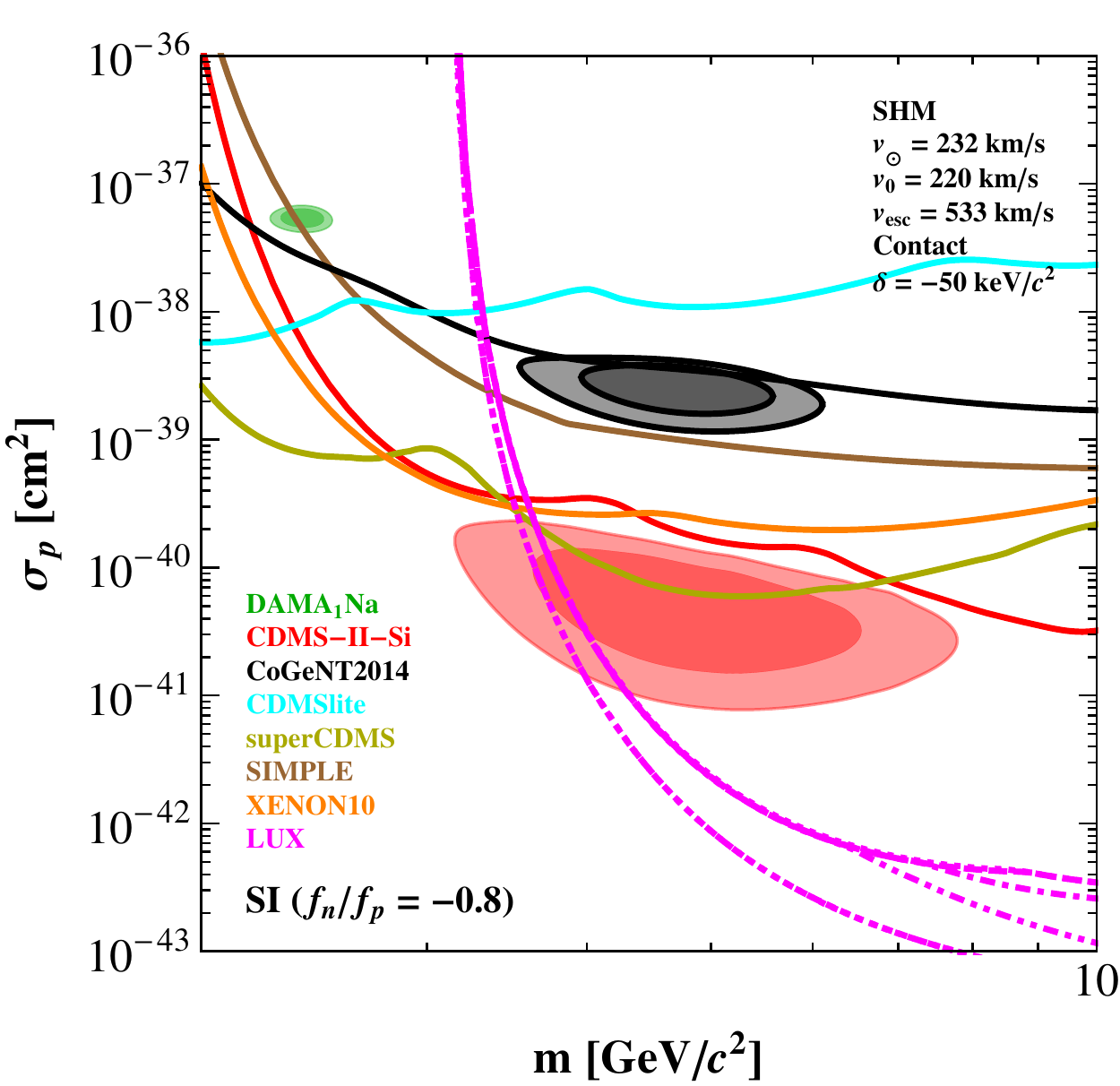}
\caption{\label{fig:fn0708delta50}
$90\%$ CL bounds and $68\%$ and $90\%$ CL allowed regions in the  WIMP-proton cross section $\sigma_p$ vs WIMP mass plane, assuming the SHM, for the spin-independent isospin-violating interactions with a) (left) $f_n/f_p =-0.7$ (``Xe-phobic")   and b) (right) $f_n/f_p =-0.8$ (``Ge-phobic"), for inelastic exothermic scattering with $\delta=-50$ keV$/c^2$.
}
\end{figure}

\begin{figure}[t]
\centering
\includegraphics[width=0.49\textwidth]{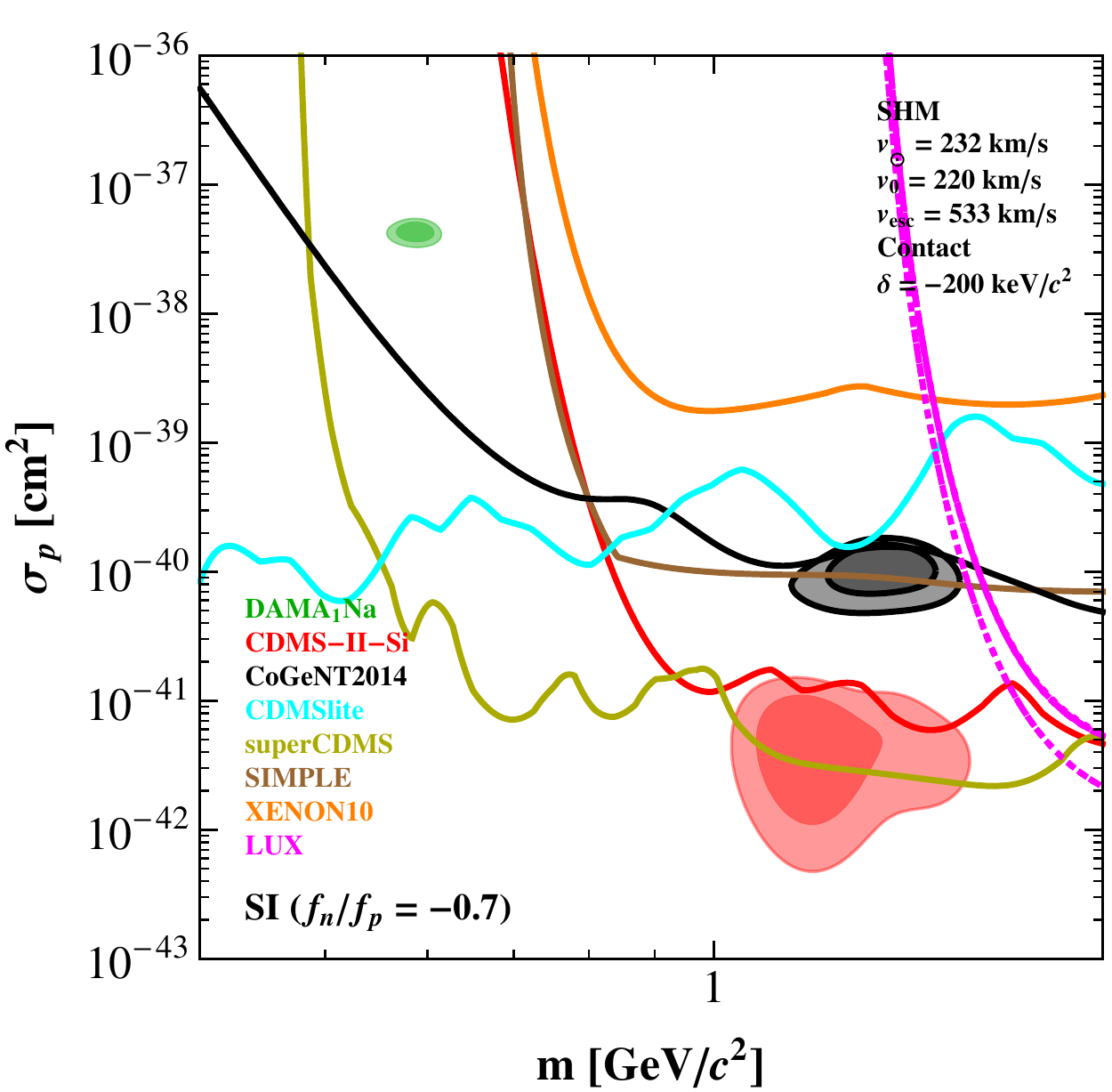}
\includegraphics[width=0.49\textwidth]{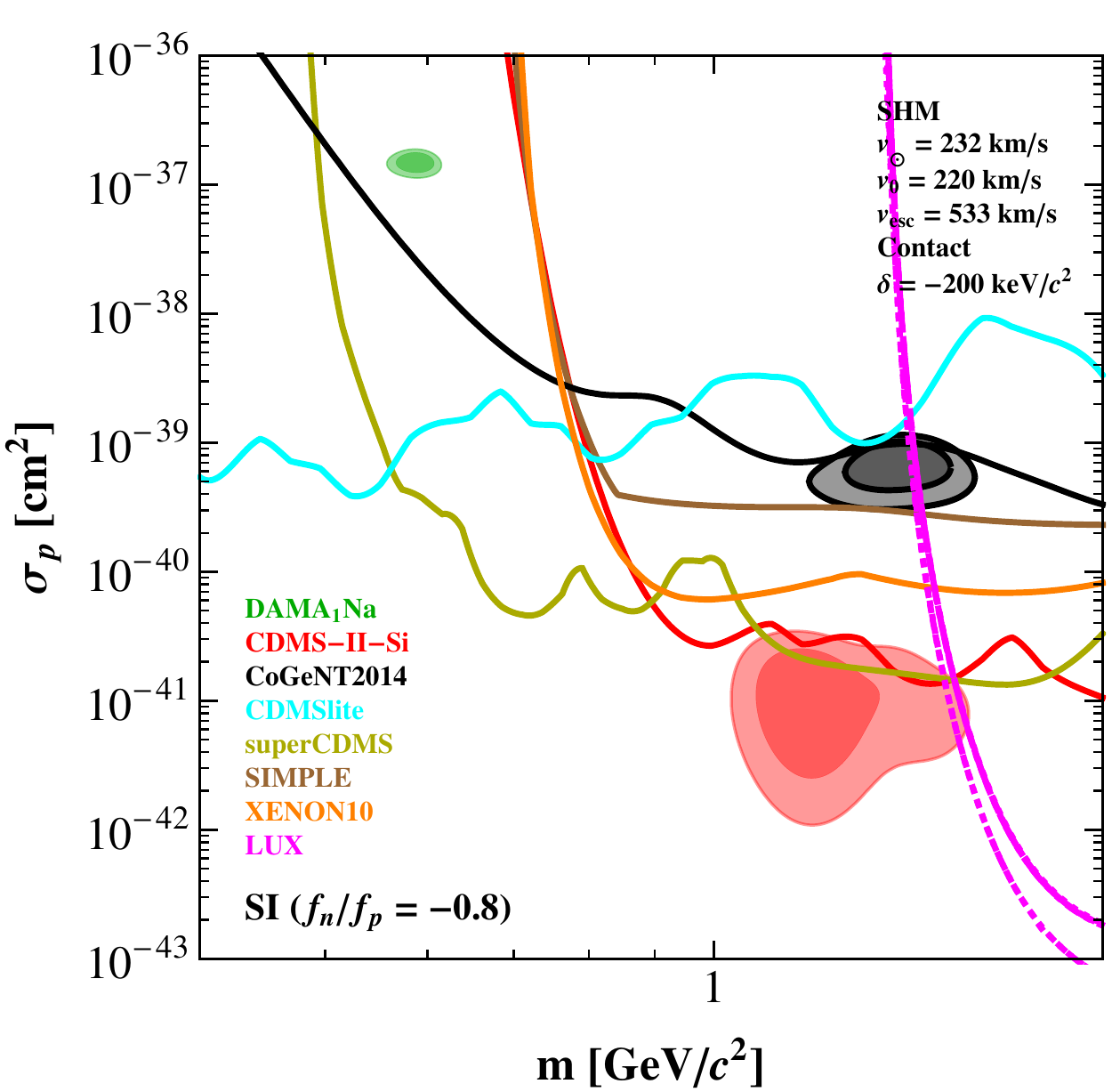}
\caption{\label{fig:fn0708delta200}
Same as Fig.~\ref{fig:fn0708delta50} but for  $\delta=-200$ keV$/c^2$.}
\end{figure}

\begin{figure}[t]
\centering
\includegraphics[width=0.49\textwidth]{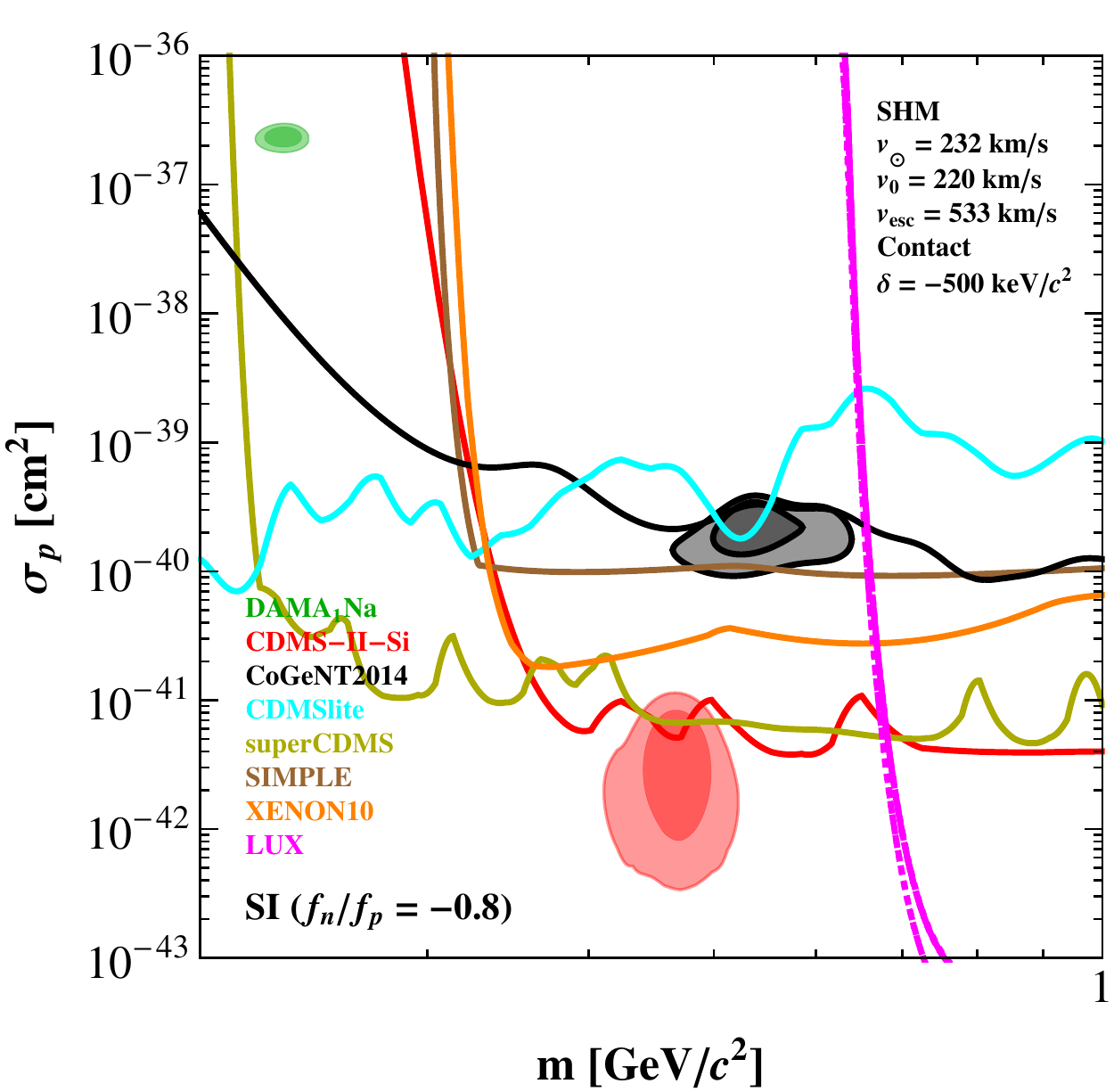}
\caption{\label{fig:fn08delta500}
$90\%$ CL bounds and $68\%$ and $90\%$ CL allowed regions in the WIMP-proton cross section $\sigma_p$ vs WIMP mass plane, assuming the SHM, for spin-independent isospin-violating interactions with $f_n/f_p =-0.8$, for inelastic exothermic scattering with $\delta=-500$ keV$/c^2$. In this case only the lowest-energy or the two lowest-energy events of the three events observed by CDMS-II-Si are due to DM.}
\end{figure}

On the other hand, as can be seen in Fig.~\ref{fig:fn08elastic}, for WIMPs with  isospin-violating  ``Ge-phobic" $f_n / f_p = -0.8$ coupling  and elastic scattering,  the 90\% CL  LUX limit rejects all  $90\%$ CL regions of interest   (although the ``Xe-phobic" coupling  $f_n / f_p = -0.7$   allows a very small sliver of the CDMS-II-Si). It is the combination of exothermic scattering (which weakens the LUX limits) and the isospin-violating couplings that could allow the CDMS-II-Si to be compatible with all present limits. This is shown in  Figs.~\ref{fig:fn0708delta50}, \ref{fig:fn0708delta200}  and  \ref{fig:fn08delta500} for $\delta= -50, -200$ and $-500$ keV/$c^2$ respectively. Notice that the isospin-violating couplings separate the CoGeNT and CDMS-II-Si regions, which instead overlap when isospin-conserving couplings are considered (see Figs.~\ref{fig:fn1elastic50} and \ref{fig:fn1delta200500}). Thus the CoGeNT region is rejected even when the 
CDMS-II-Si  is allowed.

Figs.~\ref{fig:fn0708delta50}, \ref{fig:fn0708delta200} and  \ref{fig:fn08delta500} show clearly the different effects of the ``Xe-phobic''  and ``Ge-phobic'' choices in weakening maximally the LUX and the SuperCDMS limits respectively.

\section{Halo-independent data comparison}

\begin{figure}[t]
\centering
\includegraphics[width=0.49\textwidth]{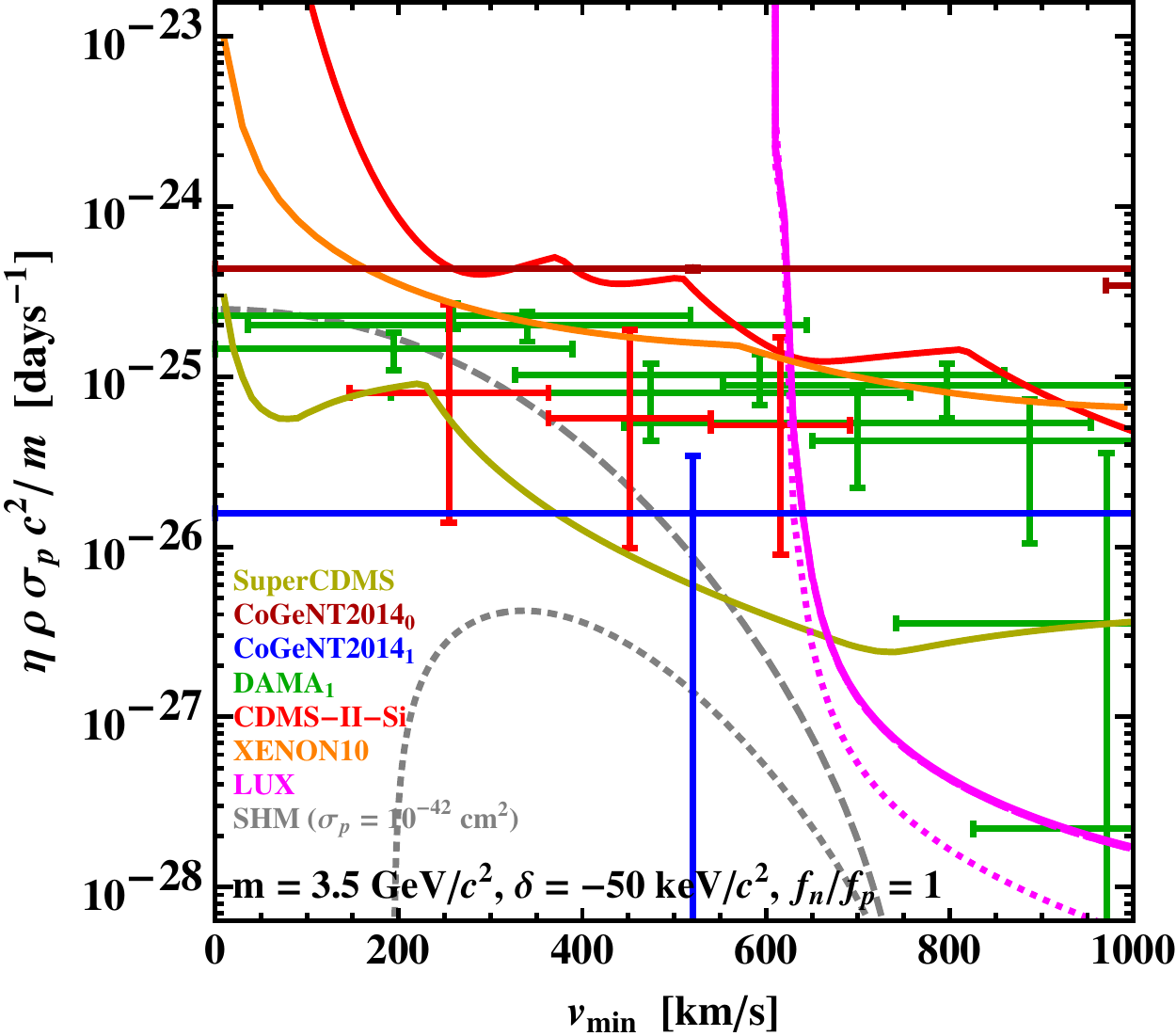}
\caption{\label{fig:HIfn1delta50}
Measurements of and upper bounds on $\tilde\eta^0(\vmin) c^2$  (for CDMS-II-Si and CoGeNT) and $\tilde\eta^1(\vmin) c^2$ (for DAMA) for inelastic exothermic scattering with  $\delta=-50$ keV$/c^2$  for a WIMP with  mass $m = 3.5$ GeV/$c^2$  and spin-independent isospin-conserving interactions.  Only the scattering in Na is considered in DAMA($Q_{\rm Na} = 0.30$). The dashed gray lines show the SHM $\tilde{\eta}^0 c^2$ (upper line) and $\tilde{\eta}^1 c^2$ (lower line) for $\sigma_p = 1 \times 10^{-42}$ cm$^2$, which in Fig.~\ref{fig:fn1elastic50}.b is within the CDMS-II-Si region. }
\end{figure}

\begin{figure}[t]
\centering
\includegraphics[width=0.49\textwidth]{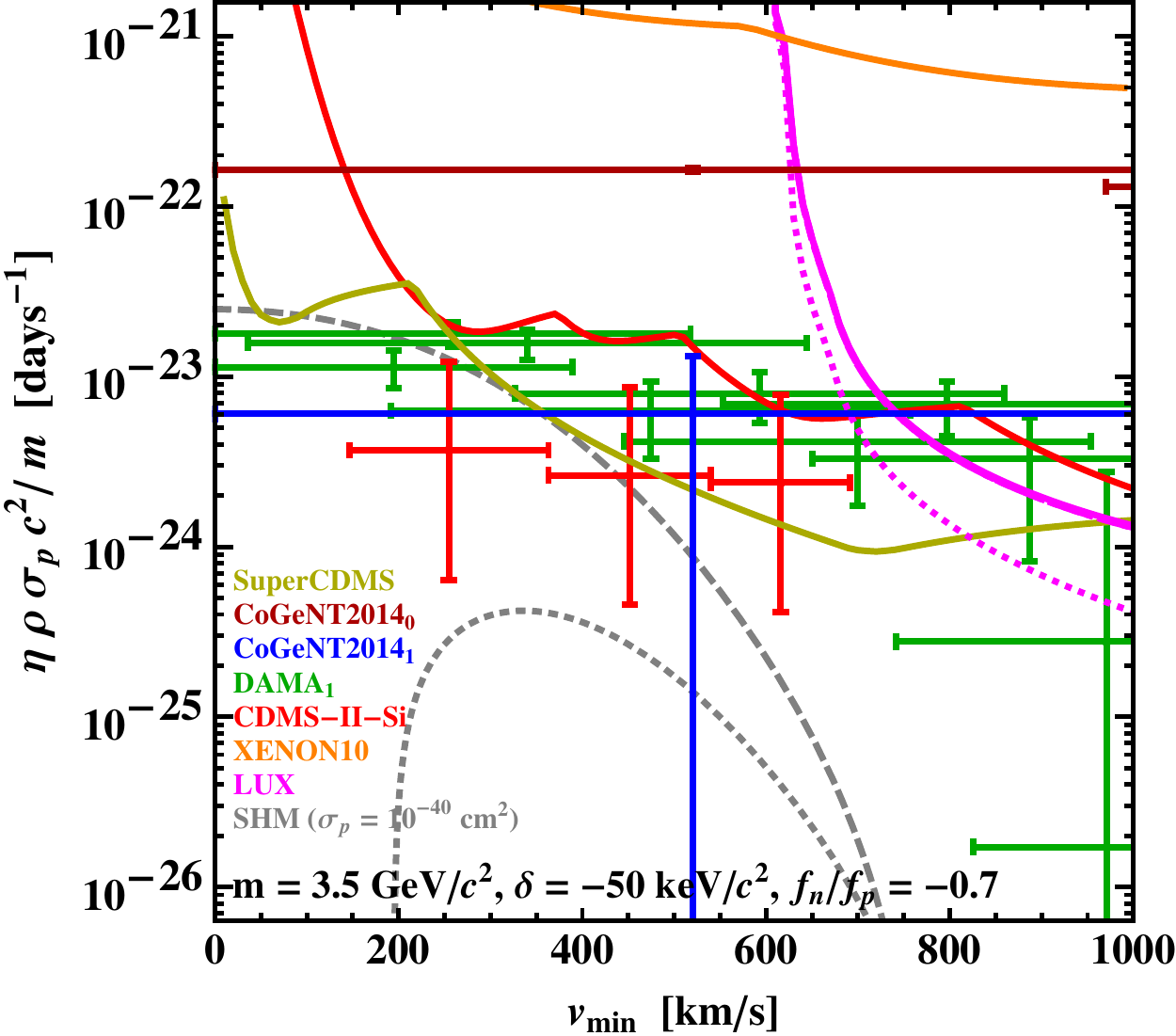}
\includegraphics[width=0.49\textwidth]{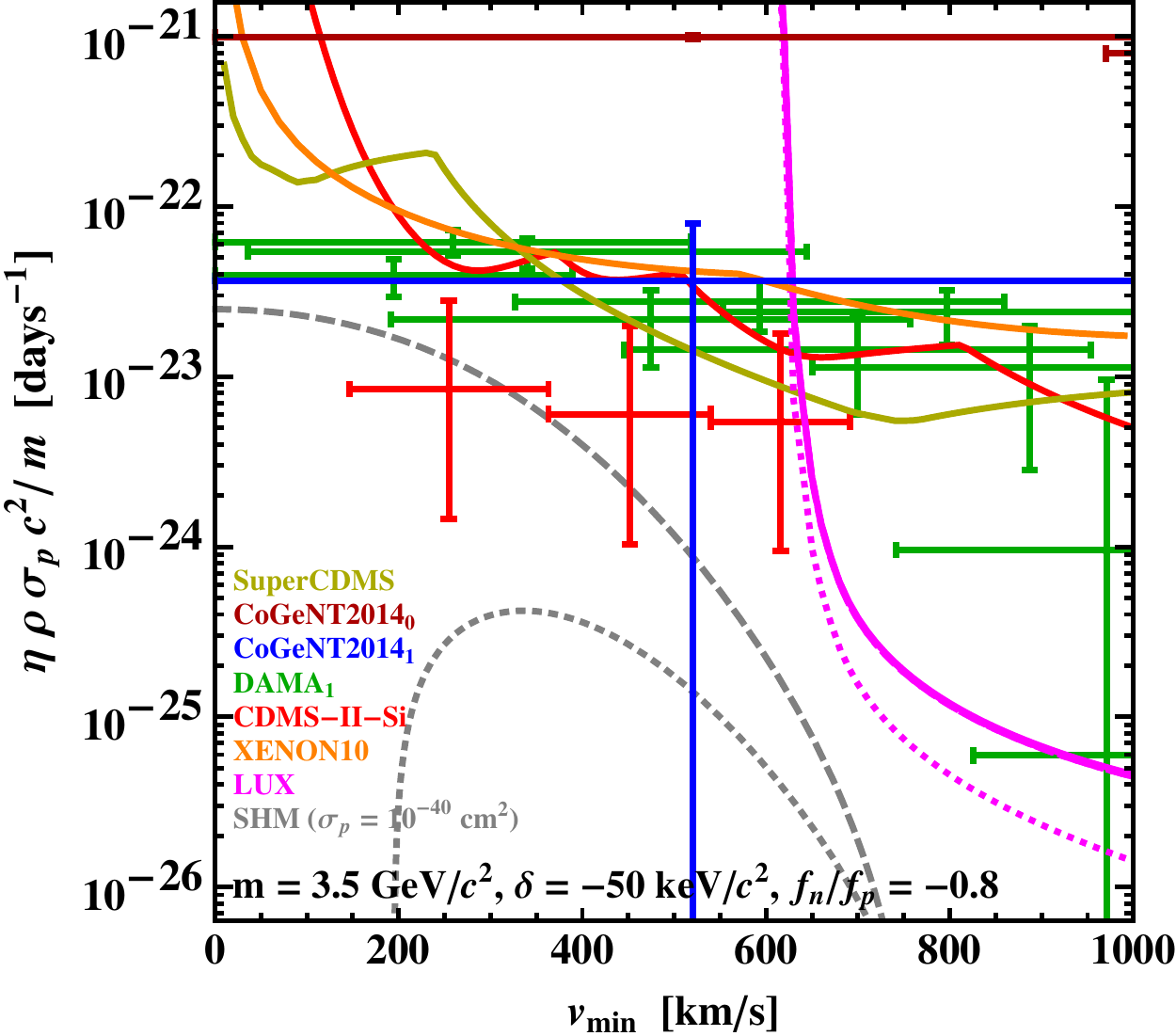}
\caption{\label{fig:HIfn0708delta50}
Same as in Fig.~\ref{fig:HIfn1delta50} but for isospin-violating  couplings with a) (left) $f_n/f_p= -0.7$ and b) (right) $f_n/f_p= -0.8$.  The dashed gray lines show the SHM $\tilde{\eta}^0 c^2$ (upper line) and $\tilde{\eta}^1 c^2$ (lower line) for $\sigma_p = 1 \times 10^{-40}$ cm$^2$, which in Fig.~\ref{fig:fn0708delta50}.b is within the CDMS-II-Si region allowed by all upper bounds.}
\end{figure}

\begin{figure}[t]
\centering
\includegraphics[width=0.49\textwidth]{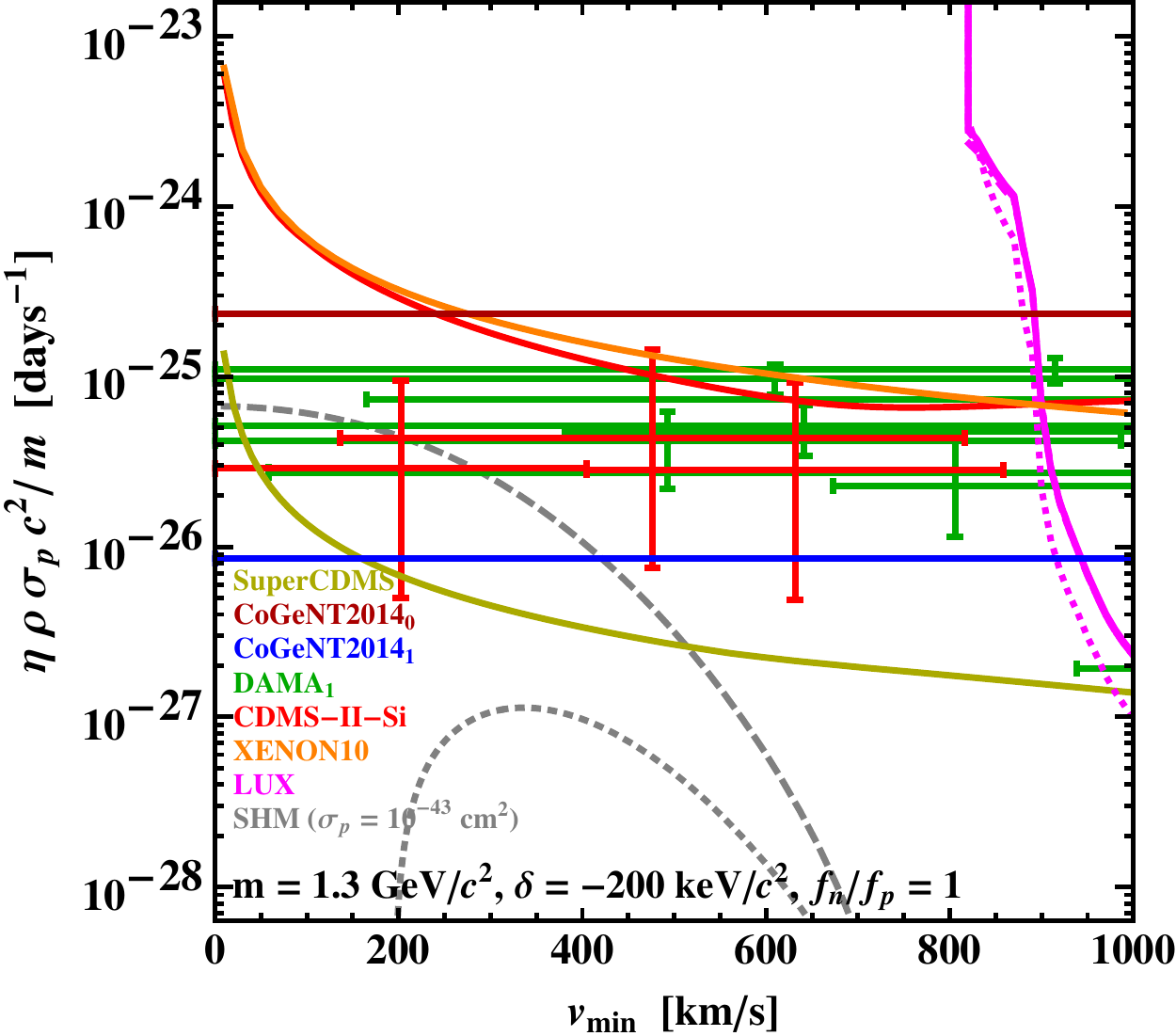}
\includegraphics[width=0.49\textwidth]{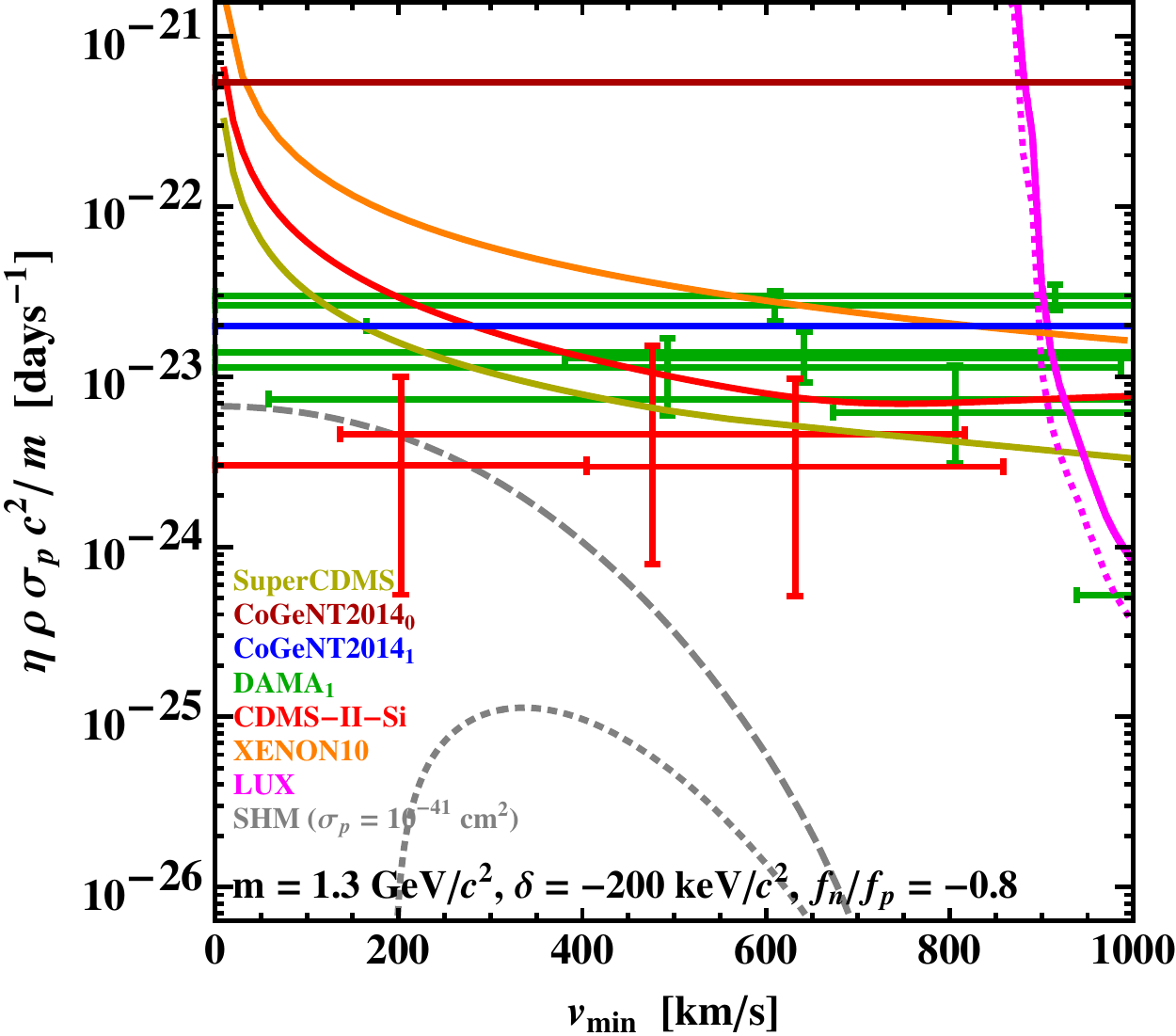}
\caption{\label{fig:HIfn108delta200}
Measurements of and upper bounds on 
$\tilde\eta^0(\vmin) c^2$  (for CDMS-II-Si and CoGeNT) and $\tilde\eta^1(\vmin) c^2$ (for DAMA) for inelastic exothermic scattering with  $\delta=-200$ keV$/c^2$  for a WIMP with mass $m = 1.3$ GeV/$c^2$  and a) (left) spin-independent isospin-conserving  interactions or b) (right) spin-independent isospin-violating coupling with $f_n/f_p= -0.8$.   The dashed gray lines show the SHM $\tilde{\eta}^0 c^2$ (upper line) and $\tilde{\eta}^1 c^2$ (lower line) for  a) $\sigma_p = 1 \times 10^{-43}$ cm$^2$ and b) $\sigma_p = 1 \times 10^{-41}$ cm$^2$ which in Fig.~\ref{fig:fn1delta200500}.a  and Fig.~\ref{fig:fn0708delta200}.b respectively are within the CDMS-II-Si region allowed by all upper bounds.
}
\end{figure}

Here we present the averages of $\tilde\eta^0(\vmin) c^2$  (for CDMS-II-Si and CoGeNT) and $\tilde\eta^1(\vmin) c^2$ (for DAMA)  compared with the most relevant upper limits, as functions of $\vmin$.   For exothermic scattering the relation between energy and $\vmin$ intervals is more complicated than for elastic scattering. Notice in Fig.~\ref{fig:ErVsV} that if the boundaries of an energy bin cross the upper $\ER$ branch, $\ER^+$,  higher recoil energies correspond to higher values of $\vmin$ (the same happens for elastic scattering, for which only the upper $E_R$ branch exists). This is the case in  Figs.~\ref{fig:HIfn1delta50} and~\ref{fig:HIfn0708delta50} for the three CDMS-II-Si  energy bins we adopted, 7 to 9 keV, 9 to 11 keV and 11 to 13 keV (each containing one observed event).  However, if the $\delta$ and $m$ values  are such that the  boundaries of an energy bin cross the lower $\ER$ branch, $\ER^-$, the $\vmin$ intervals are inverted: the largest $\vmin$ boundary  corresponds to the smallest energy  boundary and vice versa.  If instead $E_\delta$ is included in the energy interval, $\vmin$ extends all the way to $\vmin=0$. This is the case in  Fig.~\ref{fig:HIfn108delta200} for the three CDMS-II-Si  energy bins we adopted. 

Figs.~\ref{fig:HIfn1delta50} and~\ref{fig:HIfn0708delta50} show the measurements of and upper bounds on 
$\tilde\eta^0(\vmin) c^2$  (for CDMS-II-Si and CoGeNT) and $\tilde\eta^1(\vmin) c^2$ (for DAMA) for inelastic exothermic scattering with  $\delta=-50$ keV$/c^2$  for a WIMP with  mass $m = 3.5$ GeV/$c^2$. The two $E_R^\pm$ branches for this $\delta$ and $m$ combination for scattering off Si are shown as the orange lines in Fig.~\ref{fig:ErVsV}.  The CDMS-II-Si intervals in $\vmin$, shown as the horizontal bars of the three $\tilde\eta^0(\vmin) c^2$ crosses, are ordered in the same way as the three energy intervals. In Fig.~\ref{fig:HIfn1delta50}   the interaction assumed is spin-independent isospin-conserving   and the tension between the CDMS-II-Si crosses and the SuperCDMS and LUX  limits is apparent. This tension  is clearly alleviated  in Fig.~\ref{fig:HIfn0708delta50}.a and b when  the ``Xe-phobic" choice $f_n/f_p= -0.7$  or the ``Ge-phobic" choice $f_n/f_p= -0.8$ are respectively made.  This is largely the same conclusion we reached in our SHM analysis.

Our last figure, Fig.~\ref{fig:HIfn108delta200}, is more difficult to interpret than the previous ones. It corresponds to  $\delta=-200$ keV$/c^2$  and mass $m = 1.3$ GeV/$c^2$, a combination for which the two $E_R^\pm$ branches are shown in green in Fig.~\ref{fig:ErVsV}. Because our halo-independent analysis extends to larger speeds than in the SHM, up to 1000 km/s (accounting for potential extreme values of the escape velocities encountered in some halo models), the three CDMS-II-Si events are contained in the allowed recoil energy interval, as we can see in  Fig.~\ref{fig:ErVsV}.  The difficulty comes in the relation between the energy and the $\vmin$ intervals for CDMS-II-Si.  It is clear from Fig.~\ref{fig:ErVsV} that the first CDMS-II-Si energy bin, 7 to 9 keV, crosses the  lower $\ER$ branch, and  thus  its corresponding $\vmin$ interval is inverted. It is also located at higher $\vmin$ values than the interval corresponding to the second energy bin, 9 to 11 keV, which contains $E_\delta$ and thus extends to $\vmin =0$. Only the largest energy bin crosses the upper $\ER$ branch, and is as expected in elastic collisions.  It is clearly seen in Fig.~\ref{fig:HIfn108delta200}  that the SuperCDMS limit is below the CDMS-II-Si crosses in Fig.~\ref{fig:HIfn108delta200}.a, where $f_n/f_p= 1$, and it is instead above the CDMS-II-Si crosses in  Fig.~\ref{fig:HIfn108delta200}.b,  where  $f_n/f_p= -0.8$. Notice that the LUX bound in this case only affects $\vmin$ values above 800 km/s, i.e. above the maximum speed values in the SHM.

In all our halo-independent plots there is only one cross in blue and only one in brown, corresponding to the CoGeNT 2014 annual modulation and total rate respectively \cite{Aalseth:2014eft}, in the first of the two energy bins we adopted, extending from $\vmin=0$ to very large values of $\vmin$. These are almost entirely rejected by the SuperCDMS limit.

\section{Conclusions}

We have considered light WIMPs with inelastic exothermic scattering, in which a heavier DM state becomes de-excited to a lighter DM state. In our SHM analysis the CoGeNT and DAMA regions are rejected by present bounds. In our halo-independent analysis, the situation seems of strong tension, since only the lowest $\vmin$ portion of the data points remain outside the upper limits.

In both our SHM and halo-independent analyses the conclusion we reach is similar, namely that the CDMS-II-Si signal region can still be compatible with all present upper limits, in particular the LUX and the SuperCDMS limits, with a combination of two assumptions: exothermic scattering and  spin-independent isospin-violating interactions. The reason is that the exothermic character of the scattering weakens the xenon-based limits, the LUX bound in particular, but  does not weaken significantly  the SuperCDMS bound because of the low energy threshold of this experiment. This limit can be further relaxed by an isospin-violating  coupling which suppresses the WIMP-Ge coupling. In particular,  the choice of $f_n/f_p= -0.8$ for the  neutron to proton coupling ratio reduces this coupling maximally. We call this choice ``Ge-phobic". 

That nature would choose for the dark matter the particular  combination of characteristics which weakens the best experimental upper limits at a particular moment seems too much of a coincidence but, like always, more data will confirm or disprove this scenario.

\section*{Acknowledgments}

We would like to thank P. Gondolo and E. Del Nobile for useful conversations.  G.G.~and J.-H.H.~were supported in part by the Department of Energy under Award Number DE-SC0009937. J.-H.H.~was also partially supported by the Spanish Consolider-Ingenio MultiDark grant (CSD2009-00064).

\end{document}